\documentclass[preprint,superscriptaddress,amsmath,amssymb,prd,aps,showpacs,floatfix,nofootinbib]{revtex4-1}
\usepackage{graphicx}
\usepackage{epsfig,float}
\usepackage[sort&compress]{natbib}
\usepackage{subfigure}
\usepackage{amsmath}
\usepackage{amsfonts}
\usepackage{cancel}
\usepackage{lmodern,dsfont}
\usepackage{amssymb}
\usepackage{hyperref}
\usepackage{multirow}
\usepackage{rotating}
\usepackage{lipsum}
\usepackage{mathtools}
\usepackage{color}
\usepackage{bbm}
\usepackage{bbold}
\usepackage{ulem}
\newcommand{\eps}{\epsilon}

\begin{document}
\graphicspath{{./figure/}}

\title{\boldmath Molecular states of $ D^* D^* \bar{K}^*$ nature }

\author{N. Ikeno}
\email{ikeno@tottori-u.ac.jp}
\affiliation{Department of Agricultural, Life and Environmental Sciences, Tottori University, Tottori 680-8551, Japan}

\author{M. Bayar}
\email{melahat.bayar@kocaeli.edu.tr}
\affiliation{Department of Physics, Kocaeli University, 41380, Izmit, Turkey}
\affiliation{Departamento de F\'{\i}sica Te\'orica and IFIC, Centro Mixto Universidad de
Valencia-CSIC Institutos de Investigaci\'on de Paterna, Aptdo.
22085, 46071 Valencia, Spain}

\author{E. Oset}
\email{oset@ific.uv.es}
\affiliation{Departamento de F\'{\i}sica Te\'orica and IFIC, Centro Mixto Universidad de
Valencia-CSIC Institutos de Investigaci\'on de Paterna, Aptdo.
22085, 46071 Valencia, Spain}
\date{\today}

\begin{abstract}
We study the interaction of two $ D^* $ and a $\bar{K}^{*}$ 
by using the Fixed Center Approximation to the Faddeev equations to search for bound states of the three body system. Since the $ D^*  D^* $ interaction is attractive and gives a bound state, and so is the case of the $D^* \bar{K}^{*}$ interaction, where the $J^{P}=0^{+}$ bound state is identified with the $X_0 (2900)$, the $ D^*  D^* \bar{K}^{*}$  system leads to manifestly exotic bound states with $ccs$ open quarks. We obtain bound states of isospin $I=1/2$, negative parity and total spin $J=0,1,2$. For $J=0$ we obtain one state, and for $J=1,2$ we obtain two states in each case. The binding energies range from $56$ MeV to $151$ MeV and the widths from $80$ MeV to $100$ MeV.
 
 \end{abstract}

\maketitle

\section{Introduction}
While the study of general Few Body Systems has a long tradition, the study of Few Body Systems made of mesons has only a recent history. A review of such systems has been done in Ref.~\cite{albereview}. The study of the meson meson interaction with chiral Lagrangians~\cite{gasser} and the realization that this interaction, properly unitarized, led to meson--meson bound states that could be associated to known mesonic resonances~\cite{nap,ramoset,norbert,locher,juan} (see recent reviews in Refs.~\cite{ulfmale,guozou,donguo,dongzou}). The same formalism allowed one to address three body systems made with mesons, some of which could be associated to known mesonic states~\cite{albereview}.

With a few exceptions~\cite{branz}, most of the states found in the past from the meson meson interaction correspond to states which are not manifestly exotic, in the sense that they could also be formed in principle from a conventional $q \bar q$. Yet, the recent experimental findings of the $X_0 (2900)$, $X_1 (2900)$~\cite{lhcb1,lhcb2} in the $D \bar K$ invariant mass, and the $T_{cc}$ state~\cite{lhcbcc,misha} in the $D D \pi$ spectrum, revealed clear exotic mesonic structures, since one has $cs$ quarks in the first case and $cc$ quarks in the second one. These findings open the door to the formation of few body systems with several open quarks, having three of more quarks with flavors like $ccs$, $ccc$, $css$, etc., after eliminating the $q \bar q$ structures with no flavor as $u \bar u$, $d \bar d$, $s \bar s$, $c \bar c$, etc.
Since flavor is conserved in strong interactions, these systems can be relatively stable because they cannot decay into lighter systems with smaller number of mesons. In the present paper we report on calculations for the $D^* D^* \bar K^*$ system, which has open $ccs$ quarks.
The reason to choose this system is because we can establish connection with the experimental findings of Refs.~\cite{lhcb1,lhcb2,lhcbcc,misha}. Indeed, the $X_0 (2900)$ was soon identified  as a likely $D^* \bar K^*$ molecular state~\cite{xiegeng,weiwang,junhe,lugeng,qianwang,raquel,azizi,chendong,wangzhu,chensu}. Other works favor compact tetraquarks~\cite{zhigang,karliner,wangoka},
yet, the relativized quark model of Ref.~\cite{qinwang} disfavors the compact tetraquark structure and favors the molecular one.

A valuable information concerning the $D^* \bar K^*$ system comes from the work of Ref.~\cite{branz}, where 10 years prior to the observation of the $X_0 (2900)$, a $D^* \bar K^*$ molecule  had been predicted with the same quantum numbers as the $X_0 (2900)$ and a mass and width remarkably close to those of the observed ones. 
Indeed, the mass predicted in \cite{branz} for the $D^* \bar K^*$ system with $ I=0 $, $J^{P}=0^{+}$ was $2848$ MeV and the width between $23-59$ MeV. Experimentally the  $X_0 (2900)$ was found with these quantum numbers, mass $M=2866\pm 7$ MeV, and width $ \Gamma =57.2 \pm 12.9  $ MeV. A fine tuning of the parameters to obtain the exact experimental values is done in \cite{raquel}.

 The other part of the interaction is for the $D^* D^*$ subsystem. Here we rely upon the recent observation of the $T_{cc}$ state~\cite{lhcbcc,misha}. Once again the $D^* D$ nature of the $ T_{cc} $ state has been advocated from the beginning~\cite{misha} and supported by many theoretical papers~\cite{gengwang,guozou,feijoo,mehen,fisher,dengzhu,miguel,keliuli,agaev,hyodo,mengwang,abreu,chenzhang,renzhu,mengzhu,miguel2,dujuan}.
A mapping of the $D^* D$ interaction to the $D^* D^*$ system has been done in Ref.~\cite{dai} using an extrapolation of the local hidden gauge approach~\cite{hidden1,hidden2,hidden4,hideko} to the charm sector, which was already used in Ref.~\cite{branz} to study the same $D^* D^*$ exotic system.
A similar approach using the boson exchange model is done in Ref.~\cite{gengliu}. It was found in Ref.~\cite{branz} that the $D^* D^*$ system was bound in $I=0$ and $J^P = 1^+$. In Ref.~\cite{dai}, the model was refined including decay channels and using the same cut off to regularize the loops that was used in Ref.~\cite{feijoo} to describe the $T_{cc}$ state.

Three body systems of molecular nature containing two charmed quarks (or antiquarks) and a strange quark have been recently studied. In Table~1 of Ref.~\cite{albereview} one finds the $D D K$ molecule studied in Ref.~\cite{wuliugeng,alber,gengal}, $D \bar{D}^* K$ studied in Ref.~\cite{malabarba}, and $D D^* K$ studied in Ref.~\cite{mameissner}. More recently the $D \bar{D}^* K$ hexaquark state is also studied in \cite{ozdem} via QCD sum rules.
In these systems one has, however, $cc \bar s$ or $c \bar c \bar s$ quarks, but not the $ccs$ combination that we have in the system that we study, which makes it super exotic. These works are done using different technical approaches. However, for the purpose of justifying the Fixed Center Approximation (FCA) to the Faddeev equations that we follow, we find it interesting to mention two recent works on the study of the $D \bar D K$ system. In Ref.~\cite{gengwu} it is studied using the Gaussian expansion method, minimizing the energy of the system. The same system is studied in Ref.~\cite{xiangxie} using the Fixed Center Approximation (FCA) and the results obtained are very similar. The coincidence of two very different methods in a similar system to the one we study gives us confidence in the FCA method that we use in the present approach to study the molecular $D^* D^* \bar K^*$ system. The fact that we use input tuned to the $T_{cc}$ state of $D^* D$ nature in Ref.~\cite{feijoo} and the  $X_0 (2900)$ state of $D^* \bar K^*$ nature in Ref.~\cite{raquel}, gives us further confidence, not only on the existence of the bound states that we find, but also on the values of the masses and widths that we predict.

\section{\label{formalism} FORMALISM}

We will use the FCA to study the $ D^* D^* \bar{K}^{*} $ system. In this picture, there is a cluster of two bound particles and the third one collides with the components of this cluster without modifying its wave function. Certainly, if the third particle is lighter than the constituents of the cluster, the approximation is better, which suggests to consider the $ D^* D^* $ system as the cluster. This latter system has been shown to be bound in Refs. \cite{pavon,lisgengliu,dai}. In Ref. \cite{dai} the local hidden gauge approach extrapolated to the charm sector has been used, and the loops are regularized with the cut off demanded to fit the binding of the $T_{cc}$ state as a $ D^* D$ molecule in Ref. \cite{feijoo}. A state with isospin $I=0$ and $J^{P}=1^{+}$ was found. The predicted binding should be realistic and we take the results of Ref. \cite{dai} as input for the $ D^* D^* $ cluster. 

\begin{center}
  \begin{figure}
  \resizebox{0.9\textwidth}{!}{%
  \includegraphics{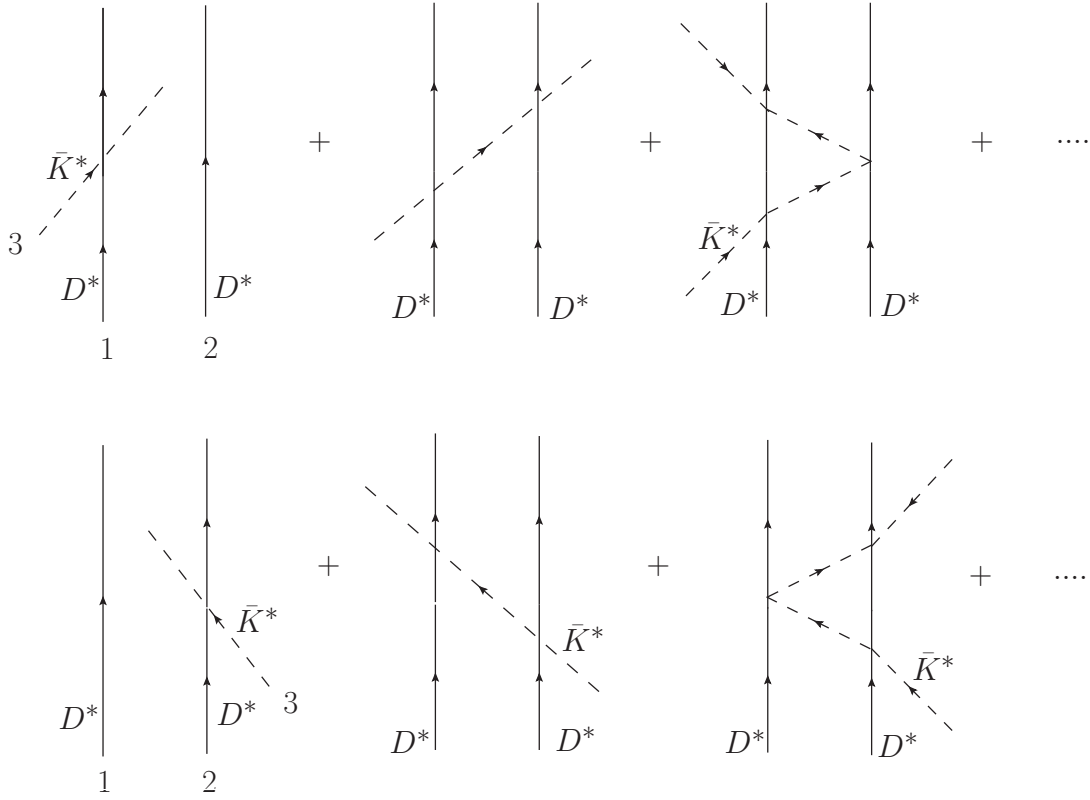}
  }
  \caption{ \label{fig:FCA}
  Diagrams involved in the FCA for the collision of the  $\bar{K}^{*}$ with the cluster of $ D^* D^* $.}
  \end{figure}
  \end{center}

We follow the formalism of Ref. \cite{luisroca}. There, the amplitude for $\bar{K}^{*}$ collision with the $D^* D^*  $  cluster is given by the sum of the partition functions $T_{1}$, $T_{2}$, where $T_{1}$ sums all diagrams in Fig. \ref{fig:FCA} where the $\bar{K}^{*}$  collides first with the particle $1$ of the cluster, while $T_{2}$ sums all diagrams where the $\bar{K}^{*}$ collides first with particle $2$ of the cluster. 
We have for the total amplitude $T$   
   \begin{eqnarray}  
   T \equiv  T_1 + T_2  \nonumber 
 \label{eq:TotalT} 
  \end{eqnarray}
and $T_{1}$, $T_{2}$ are coupled through 
   \begin{eqnarray}  
  T_1 &=&t_1 + t_1 G_0 T_2, \nonumber \\
  T_2&=&t_2 + t_2 G_0 T_1, \nonumber 
  \label{eq:T1T2} 
  \end{eqnarray}
where $t_{1}$ is the scattering amplitude for $D^*(1) \bar{K}^{*} $, and $t_{2}$ the corresponding amplitude for $D^*(2) \bar{K}^{*} $  scattering, and $G_0$ is the $\bar{K}^{*} $ propagator folded with the cluster wave function.  However, we should take into account the isospin and spin of the $  D^*(i) \bar{K}^{*}$ amplitudes, where $i$ refers to any  $  D^* $ of the cluster.

\subsection{\label{isospin} ISOSPIN CONSIDERATIONS}

With the isospin doublet $(D^{+}, - D^{0})$, the $I=0$ $D^* D^*$ state is given by 

\begin{eqnarray}
  \mid D^{*} D^{*}, ~I=0 \rangle =-\frac{1}{\sqrt{2}} ( D^{*+} D^{*0}- D^{*0} D^{*+}) \nonumber \\
\end{eqnarray}
and one should also bear in mind that this is accompanied by $ J=1 $ in each of the two components, which we will address below. 

When evaluating $t_{1}$ of the scattering amplitude for $D^*(1) \bar{K}^{*}$, we have the following matrix element using the notation
$ \mid I_{3} (D^*(1)),I_{3} (D^*(2))\rangle \mid I_{3} (\bar{K}^{*})\rangle$, where we have chosen $\bar{K}^{*} $ to have the $ I_{3}=\frac{1}{2} $ component
\begin{eqnarray}  
 \frac{1}{\sqrt{2}} \frac{1}{\sqrt{2}} && \left(  \langle D^* D^*, \frac{1}{2}, -\frac{1}{2} \mid - \langle D^* D^*, -\frac{1}{2}, \frac{1}{2} \mid \right) \nonumber \\
  && \times \langle \bar{K}^{*}, \frac{1}{2} \mid  \mid t_1 \mid  \left(  \mid D^* D^*, \frac{1}{2}, -\frac{1}{2} \rangle - \mid D^* D^*, -\frac{1}{2}, \frac{1}{2} \rangle \right)  \mid \bar{K}^{*}, \frac{1}{2} \rangle.  
   \label{IsoXX}
\end{eqnarray}
To make connection with the $  D^* \bar{K}^{*}  $ isospin amplitudes of Refs. \cite{branz,raquel}, we combine the third component of $ D^*(1) $ with the one of the $ \bar{K}^{*}  $ to give states of $  D^* \bar{K}^{*}  $  isospin. Hence, we have with the notation  
$ \mid  I (D^*(1) \bar{K}^{*}),~  I_{3} (D^*(1) \bar{K}^{*}) \rangle \mid  I_{3} (D^{*}(2))  \rangle $ 
\begin{eqnarray}
 &&\dfrac{1}{2} \left[   \left( \langle 1,1 \mid \langle - \frac{1}{2}  \mid -\dfrac{1}{\sqrt{2}}  (\langle 1,0 \mid - \langle 0,0 \mid ) \langle \frac{1}{2}     \mid  \right)  \vert~t_1 ~ \vert \left(  \mid 1,1  \rangle  \mid - \frac{1}{2} \rangle - \dfrac{1}{\sqrt{2}} ( \mid 1 0 \rangle  -\mid 0 0\rangle)  \mid \frac{1}{2} \rangle \right) \right] \nonumber \\
 &&= \dfrac{1}{2} \left( t^{I=1} + \dfrac{1}{2} t^{I=1} +  \dfrac{1}{2} t^{I=0} \right) =  \dfrac{3}{4} t^{I=1}+ \dfrac{1}{4} t^{I=0},
  \label{IsospinTot}
\end{eqnarray} 
where the final result stems since $t_1$ only affects  $  D^*(1) \bar{K}^{*}  $, considering that $ D^*(2)$ is a spectator and has to be the same in the bra and the ket of the matrix element. The amplitude $t_2$, for $D^*(2)\bar{K}^{*}$ scattering is obviously equal to $t_1$.

\subsection{\label{spin} SPIN CONSIDERATIONS} 

We start with $ D^* D^* $ with $ J=1 $ and the $ \bar{K}^{*} $ has also $ J=1 $. Hence, we can have three total spins for $ D^* D^* \bar{K}^{*} $, $ J=0,1,2 $. Let us see each one of the cases. 

\subsubsection{\label{spin0} TOTAL $J=0$}

With the notation $ \mid  j_{3} (D^* D^* ), j_{3} (\bar{K}^{*}) \rangle $,  we have 
\begin{equation}
 \mid D^* D^* \bar{K}^{*}, J=0  \rangle =\dfrac{1}{\sqrt{3}} \left(  \mid 1,-1  \rangle - \mid 0,0  \rangle + \mid -1,1  \rangle \right)   \nonumber
\end{equation}
and now we write the state $ \mid  j (D^* D^* )=1 , j_{3} (D^* D^* ) \rangle $ in terms of the $  \mid j_{3} (D^*(1)) j_{3} (D^*(2))  \rangle$ and we have 
\begin{eqnarray}  
\dfrac{1}{\sqrt{3}} \left\lbrace \dfrac{1}{\sqrt{2}} (\mid 1,0  \rangle - \mid 0,1  \rangle) \mid -1  \rangle - \dfrac{1}{\sqrt{2}} (\mid 1,-1  \rangle - \mid -1,1  \rangle) \mid 0 \rangle + \dfrac{1}{\sqrt{2}} (\mid 0,-1  \rangle - \mid -1,0  \rangle) \mid 1 \rangle \right\rbrace  \nonumber
\end{eqnarray}
And now we write $ \mid j_{3} (D^*(1)) j_{3} (\bar{K}^{*}) \rangle$ in terms of $ \mid j(D^* \bar{K}^{*}) j_{3} (D^* \bar{K}^{*}) \rangle $, 
and with the notation $ \mid  j(D^*(1) \bar{K}^{*}) , j_{3} (D^*(1) \bar{K}^{*}) \rangle  \mid j_{3} (D^*(2)) \rangle  $, we have
\begin{eqnarray}  
&& \dfrac{1}{\sqrt{3}}  \dfrac{1}{\sqrt{2}} \left\lbrace  \left( \dfrac{1}{\sqrt{6}} \mid 2,0  \rangle  + \dfrac{1}{\sqrt{2}}  \mid 1,0  \rangle +  \dfrac{1}{\sqrt{3}} \mid 0,0  \rangle \right)  \mid 0  \rangle    - \left( \dfrac{1}{\sqrt{2}}  \mid 2,-1  \rangle  + \dfrac{1}{\sqrt{2}}  \mid 1,-1  \rangle \right) \mid 1 \rangle \right.  \nonumber \\ 
&& - \left( \dfrac{1}{\sqrt{2}}  \mid 2,1  \rangle  + \dfrac{1}{\sqrt{2}}  \mid 1,1  \rangle \right) \mid -1 \rangle + \dfrac{1}{\sqrt{2}} \left( \mid 2,-1  \rangle - \mid 1,-1  \rangle \right)   \mid 1  \rangle \nonumber \\
&& \left.   +\dfrac{1}{\sqrt{2}} \left( \mid 2,1  \rangle - \mid 1,1  \rangle \right)   \mid -1  \rangle - \left( \dfrac{1}{\sqrt{6}} \mid 2,0  \rangle - \dfrac{1}{\sqrt{2}}  \mid 1,0  \rangle +  \dfrac{1}{\sqrt{3}} \mid 0,0  \rangle \right) \mid 0  \rangle \right\rbrace \nonumber \\ 
&&= \dfrac{1}{\sqrt{3}}  \left\lbrace  \mid 1,0  \rangle \mid 0  \rangle -\mid 1,-1 \rangle \mid 1  \rangle - \mid 1,1  \rangle \mid -1  \rangle \right\rbrace   \nonumber
\end{eqnarray}
and the matrix element of $ t_1 $ sandwiched between this state, considering that the $ D^*(2) $ state is a spectator, becomes 
\begin{eqnarray}
 t_1 = \dfrac{1}{3}  \left( t^{j=1}+ t^{j=1} +t^{j=1} \right) = t^{j=1}. \nonumber
\end{eqnarray}
The result is expected since if we have $ \mid (D^*(1) \bar{K}^{*} ) D^*(2) \rangle $ with $j(D^*(2)) =1$, the spin of $  D^*(1)\bar{K}^{*}$ must necessarily  be $1$ to match total spin zero for the $ (D^*(1) \bar{K}^{*} ) D^*(2) $ system.

\subsubsection{\label{spin0} TOTAL $J=1$,  $ J_{3}=1 $ }
The result cannot depend on the third component and we choose it to be $ J_{3} =1 $ for simplicity.
Once again, with the $ \mid  j_{3} (D^* D^* ), j_{3} (\bar{K}^{*}) \rangle $ representation we have the state 
\begin{equation}
  \dfrac{1}{\sqrt{2}} \left\lbrace \mid 1,0  \rangle - \mid 0,1  \rangle \right\rbrace \nonumber
\end{equation}
which in the $ \mid  j_{3} (D^*(1)) j_{3} (D^*(2))  \rangle \mid  j_{3} (\bar{K}^{*})  \rangle $ representation reads 
\begin{eqnarray}  
\dfrac{1}{\sqrt{2}} \left\lbrace \left(  \dfrac{1}{\sqrt{2}} \mid 1,0  \rangle - \dfrac{1}{\sqrt{2}} \mid 0,1  \rangle \right)  \mid 0 \rangle - \left(  \dfrac{1}{\sqrt{2}} \mid 1,-1  \rangle - \dfrac{1}{\sqrt{2}} \mid -1,1  \rangle \right)  \mid 1 \rangle \right\rbrace  \nonumber
\end{eqnarray}
and written in terms of  $ \mid  j(D^* (1) \bar{K}^{*}) j_{3} (D^* (1) \bar{K}^{*})  \rangle  \mid  j_{3} (D^* (2) ) \rangle  $ is 
\begin{eqnarray}  
&&   \dfrac{1}{2} \left\lbrace  \left( \dfrac{1}{\sqrt{2}} \mid 2,1  \rangle  + \dfrac{1}{\sqrt{2}}  \mid 1,1  \rangle  \right)  \mid 0  \rangle    + \left( -\sqrt{\dfrac{2}{3}}  \mid 2,0  \rangle  + \dfrac{1}{\sqrt{3}}  \mid 0,0 \rangle \right) \mid 1 \rangle \right.  \nonumber \\  
&&\left.  -  \mid 2,2  \rangle  \mid -1  \rangle + \left( \dfrac{1}{\sqrt{6}}  \mid 2,0  \rangle  - \dfrac{1}{\sqrt{2}}  \mid 1,0  \rangle + \dfrac{1}{\sqrt{3}}  \mid 0,0  \rangle \right) \mid 1 \rangle  \right\rbrace  \nonumber \\
&&=  \dfrac{1}{2} \left\lbrace  \left( \dfrac{1}{\sqrt{2}} \mid 2,1  \rangle  + \dfrac{1}{\sqrt{2}}  \mid 1,1  \rangle \right) \mid 0 \rangle - \dfrac{1}{\sqrt{6}} \mid 2,0  \rangle  \mid 1 \rangle +\dfrac{2}{\sqrt{3}} \mid 0,0  \rangle  \mid 1 \rangle \right.  \nonumber \\
&& \left. - \mid 2,2  \rangle  \mid -1 \rangle -\dfrac{1}{\sqrt{2}} \mid 1,0  \rangle  \mid 1 \rangle
 \right\rbrace .  \nonumber
\end{eqnarray}
The $ t_1 $ matrix sandwiched between this state gives the combination 
\begin{eqnarray}
 t_1 = \dfrac{1}{4}  \left(\dfrac{5}{3}  t^{j=2}+ t^{j=1} + \dfrac{4}{3} t^{j=0} \right).  \nonumber
\end{eqnarray}

\subsubsection{\label{spin0} TOTAL $J=2$,  $ J_{3}=2 $ }

Once again the result does not depend on the third component, which we choose to be $ J_{3} =2 $.
In the $ \mid  j_{3} (D^* D^* ), j_{3} (\bar{K}^{*}) \rangle $  representation the state is
\begin{equation}  
  \mid 1,1 \rangle.  \nonumber 
  \end{equation}
Then with the $ \mid j_{3} (D^*(1)) j_{3} (D^*(2)) \rangle \mid j_{3} (\bar{K}^{*}) \rangle $ notation  we have
\begin{eqnarray}  
 \left(  \dfrac{1}{\sqrt{2}} \mid 1,0  \rangle - \dfrac{1}{\sqrt{2}} \mid 0,1  \rangle \right)  \mid 1 \rangle  \nonumber
\end{eqnarray}
which in the  $ \mid  j(D^* (1) \bar{K}^{*}) j_{3} (D^* (1) \bar{K}^{*}) \rangle \mid j_{3} (D^* (2) ) \rangle $ representation  reads
\begin{eqnarray}  
 \dfrac{1}{\sqrt{2}}  \mid 2,2  \rangle  \mid 0 \rangle -  \dfrac{1}{\sqrt{2}} \left(  \dfrac{1}{\sqrt{2}} \mid 2,1 \rangle - \dfrac{1}{\sqrt{2}} \mid 1,1  \rangle \right)  \mid 1 \rangle.  \nonumber
\end{eqnarray}
The $ t_1 $ matrix sandwiched between this state is then
\begin{eqnarray}
 t_1 = \dfrac{3}{4}  t^{j=2} +  \dfrac{1}{4} t^{j=1}. \nonumber
\end{eqnarray}

\subsubsection{\label{COMBINESPIN} COMBINED SPIN AND ISOSPIN AMPLITUDE  }
Combining the isospin and the spin decomposition of the amplitudes in subsections \ref{isospin} and \ref{spin}, we find the final contributions

\begin{eqnarray}
 t_1 &=& \dfrac{3}{4}  t^{I=1,~j=1} +  \dfrac{1}{4} t^{I=0,~j=1}, ~~~\text{for} ~~~J=0 \nonumber \\
 t_1 &=& \dfrac{1}{16} \left\lbrace 5 t^{I=1,~j=2}+ 3 t^{I=1,~j=1}+ 4 t^{I=1,~j=0}+ \dfrac{5}{3} t^{I=0,~j=2}\right. \nonumber \\
  &+& \left. t^{I=0,~j=1}+\dfrac{4}{3} t^{I=0,~j=0} \right\rbrace, ~~~\text{for} ~~~J=1\nonumber \\
  t_1 &=& \dfrac{1}{16} \left\lbrace 9 t^{I=1,~j=2}+ 3 t^{I=1,~j=1}+ 3 t^{I=0,~j=2}+  t^{I=0,~j=1} \right\rbrace, ~~~\text{for} ~~~J=2 .
  \label{eq:AmpJ012}
\end{eqnarray}

\subsection{\label{normalization } NORMALIZATION OF THE AMPLITUDES} 

By looking at the papers \cite{junko,bayar} and the diagram of double scattering in Fig. \ref{fig:FCA}, we find for the $ S $ matrix

   \begin{eqnarray}
  S^{(2)}&=&
  -i  (2\pi)^4  \delta^4 (p_{fin}-p_{in})   \frac{1}{ {\cal V}^2 } \nonumber \\
  &\times& 
  \frac{1}{ \sqrt{2\omega_{\bar{K}^{*}}} } \frac{1}{ \sqrt{2\omega_{\bar{K}^{*}}} }
  \frac{1}{ \sqrt{2\omega_{D^*}} } \frac{1}{ \sqrt{2\omega_{D^*}} }
  \frac{1}{ \sqrt{2\omega_{D^*}} } \frac{1}{ \sqrt{2\omega_{D^*}} } \nonumber \\
  &\times& 
  t_{1} t_{2} \int \frac{d^3 q}{(2\pi)^3} F(\vec{q}) \frac{1}{ q^{0^{2}} -\vec{q}~^2 - m^{2}_{{\bar{K}^{*}}}  + i \epsilon }   ,
  \label{eq:S2a}
  \end{eqnarray}
where $ \cal V $ is the volume of the normalization box,  $  p_{fin}$ and $p_{in} $ the final and initial momenta of the three particle system, $ q $ is the momentum of the exchanged $ \bar{K}^{*} $, $w_{i}=\sqrt{ m^{2}_{i}+\vec{q}_{i}^{~2}}$ and $  F(\vec{q}) $ the form factor of the cluster. If we look at the process from the macroscopic perspective of having $  (D^*(1) D^*(2))_{C} \bar{K}^{*}$, with $(D^*(1) D^*(2))_{C} $ meaning the $  D^* D^* $ cluster, the $S$ matrix reads

   \begin{eqnarray}
  S^{(2)}=
  -i  (2\pi)^4  \delta^4 (p_{fin}-p_{in})   \frac{1}{ {\cal V}^2 } 
  \frac{1}{ \sqrt{2\omega_{\bar{K}^{*}}} } \frac{1}{ \sqrt{2\omega_{\bar{K}^{*}}} }
  \frac{1}{ \sqrt{2\omega_{C}} } \frac{1}{ \sqrt{2\omega_{C}} } T^{(2)}  ,
  \label{eq:S2}
  \end{eqnarray}
Hence
 \begin{eqnarray}
  T^{(2)}&=&2\omega_{C} ~t_{1} t_{2} ~\frac{1}{ 2\omega_{D^*} } \frac{1}{ 2\omega_{D^*} }
 \int \frac{d^3 q}{(2\pi)^3} F(\vec{q}) \frac{1}{ q^{0^{2}} -\vec{q}~^2 - m^{2}_{{\bar{K}^{*}}}  + i \epsilon }  \nonumber \\
  &=& \dfrac{2\omega_{C}}{2\omega_{D^*}} ~  \dfrac{2\omega_{C}}{2\omega_{D^*}} ~ \dfrac{1}{2\omega_{C}} 
  ~t_{1} t_{2} \int \frac{d^3 q}{(2\pi)^3} F(\vec{q}) \frac{1}{ q^{0^{2}} -\vec{q}~^2 - m^{2}_{{\bar{K}^{*}}}  + i \epsilon }. 
  \label{eq:T2}
  \end{eqnarray}
 It is  thus convenient to write the partition functions suited to the macroscopic formalism as 
 
  \begin{equation}  
  \tilde{T_1} =\tilde{t_1} + \tilde{t_1} \tilde{G_0}  \tilde{T_2};~~~~   \tilde{T_2}=\tilde{t_2} + \tilde{t_2} \tilde{G_0} \tilde{T_1}, 
  \label{eq:tildeT1T2} 
  \end{equation}
where approximating $ \omega_{D^*} = m_{D^*} $, $ \omega_{C} = m_{C} $
  \begin{equation}
   \tilde{t_1}=\frac{2m_C}{2m_{D^*}}~ t_1;~~~~\tilde{t_2}=\frac{2m_C}{2m_{D^*}}~ t_2, \nonumber
    \label{eq:oran}
  \end{equation}
 and 
 \begin{equation}
\tilde{G_0}=\frac{1}{2 m_{C}}\int\frac{d^3q}{(2\pi)^3}F(\vec{q})\frac{1}{{q^0}^2-\vec{q}\,^2-m_{\bar{K}^{*}}^2+i\epsilon}.
\label{Eq:G0}
\end{equation}
The wave function of the cluster enters through $ \tilde{G_0} $ via the form factor $ F( \vec{q}) $. The function $ \tilde{G_0} $  corresponds to the propagator of $ K^* $ between two scatterings in Fig. \ref{fig:FCA}, folded with the wave function of the cluster. 

The form factor is $ F(q) = \int d^{3}r e^{i \vec{q}\cdot \vec{r} } \mid \Psi(\vec{r}) \mid^{2} $, but we find convenient to write it in terms of the wave function written in momentum space. For this purpose let us recall that the unitary approach that we use to obtain the $  D^* D^* $ bound states can be easily visualized as coming from the use of a separable potential of the type 
 \begin{equation}
V(\vec{q}, \vec{q}~')= V ~\theta(q_{max}-\mid \vec{q} \mid) ~ \theta(q_{max}-\mid \vec{q}~' \mid)
\label{Eq:Vvv}
\end{equation}
from where one can easily deduce the wave function in momentum space as \cite{danijuan}
\begin{equation}
 \Psi(\vec{p})= g_{R}~\dfrac{\theta(q_{max}-\mid \vec{p} \mid)}{E-\omega_{1}(\vec{p})-\omega_{2}(\vec{p})}
\label{Eq:Psip}
\end{equation}
where $ g_{R} $ is the coupling of the state to the two components of the state and $ \omega_{i}= \sqrt{m^{2}_{i}+\vec{p}^{~2}}$. Then, using the Fourier transform of $ \Psi(\vec{q}) $ to go to $ \Psi(\vec{r}) $, one immediately finds the form factor $  F(\vec{q}) $, normalized as $  F(\vec{q}=0)=1 $, written in terms of $ \Psi(\vec{p}) $ as \cite{luisroca, junko},

\begin{eqnarray}
  F(q) &=& \frac{1}{{\cal N}}
  \int_{ \stackrel{p < q_{{ max} }}{|{\vec{p} - \vec{q}}| < q_{{max}}}}
  d^3p ~   
  ~ \frac{1}{m_{C}-\sqrt{m^{2}_{D^*}+ \vec{p}^{~2}} -\sqrt{m^{2}_{D^*}+ \vec{p}^{~2}}} \nonumber \\
 &\times& 
   \frac{1}{m_{C}-\sqrt{m^{2}_{D^*}+ (\vec{p}-\vec{q})^{2}} -\sqrt{m^{2}_{D^*}+ (\vec{p}-\vec{q})^{2}}},
   \label{eq:formfactor}
  \end{eqnarray}
with the normalization constant ${\cal N}$
  \begin{eqnarray}
  {{\cal N }}
  =
  \int_{p < q_{{max}}} d^3p \left( ~ \frac{1}{m_{C}-\sqrt{m^{2}_{D^*}+ \vec{p}^{~2}} -\sqrt{m^{2}_{D^*}+ \vec{p}^{~2}}} \right)^2  \nonumber 
  \label{eq:normformf}
  \end{eqnarray}

With the potential of Eq. (\ref{Eq:Vvv}) the $t$ matrix is also separable 
 \begin{equation}
T(\vec{q}, \vec{q}~')= \theta(q_{max}-\mid \vec{q} \mid) ~ \theta(q_{max}-\mid \vec{q}~' \mid) ~t
\label{Eq:tseperable}
\end{equation}
with
%
\begin{equation}
t=V+VG_{D^* D^*} t, 
\label{Eq:BetheS}
\end{equation}
and $ G_{D^* D^*} $, the $ D^* D^* $ loop function, is given by 
 \begin{equation}
G_{D^* D^*}(M_{inv})= \int_{\mid q \mid < q_{{max}}} \frac{d^3q}{(2\pi)^3} \dfrac{\omega_{1}(q)+\omega_{2}(q)}{2\omega_{1}(q) \omega_{2}(q)} \dfrac{1}{M^{2}_{inv}-(\omega_{1}(q)+\omega_{2}(q))^{2}+i~\epsilon}.
\label{Eq:G0loopf}
\end{equation}
For the value of $ q_{max} $  we use $  q_{max}=420 $ MeV, which was used in Ref. \cite{dai}. This value is the one needed to get the $ T_{cc} $ state in Ref. \cite{feijoo}. With this representation of the wave function, $ F(\vec{q}) $ vanishes for $ |{\vec{q}}| > 2~q_{{max}} $. 

Following Refs. \cite{luisroca, junko} we take $ q^{0} $ in the rest frame of the cluster, which in the present case is 
\begin{equation}
q^0=\frac{s-m_{\bar{K}^{*}}^2-m_{C}^2}{2 m_{C}} \nonumber  
\end{equation}
with $ s $ the square of the total rest energy of the $ (D^* D^* )_{C} \bar{K}^{*} $ system. We also need the argument of the  $ t_1 $ matrix of the  $ D^*  \bar{K}^{*}   $ subsystem. This is evaluated taking 
\begin{eqnarray}
s^{'}&=&(p_{\bar{K}^{*}}+\dfrac{1}{2} P_{C})^{2} = m_{\bar{K}^{*}}^{2} + \dfrac{1}{4} m_{C}^{2} + q^0 m_{C}\nonumber \\
&=&\dfrac{1}{2} \left( s-m_{\bar{K}^{*}}^{2} - m_{C}^{2} \right) +\dfrac{1}{4} m_{C}^{2} + m_{\bar{K}^{*}}^{2}
\simeq  \dfrac{1}{2} \left( s-m_{\bar{K}^{*}}^{2} - m_{C}^{2} \right) + m_{D^*}^{2} + m_{\bar{K}^{*}}^{2},\nonumber 
\end{eqnarray}
the last result corresponding to the formula used in Refs. \cite{luisroca, junko}.

We  should note that, since $ t_1 = t_2 $, then  $ T_1 = T_2 $ and, thus, Eq. (\ref{eq:tildeT1T2} ) leads to 
\begin{equation}  
  \tilde{T_1} =\tilde{t_1} + \tilde{t_1} \tilde{G_0}  \tilde{T_1};~~~~   \tilde{T_1}=\dfrac{1}{\tilde{t_1}^{-1}-\tilde{G_0}  }; ~~ \tilde{T}= \tilde{T_1}+ \tilde{T_2}=2 \tilde{T_1}
  \label{eq:tildeTtotal} 
  \end{equation}
and we shall plot $ \mid \tilde{T} \mid^{2} $ for real values of $s$, looking for the peaks, from where we deduce the mass and width of the states that we find.

\subsection{\label{ELEMENTARYamp } ELEMENTARY AMPLITUDES} 

We need the $ D^* \bar{K}^{*} $ amplitudes for the different $ I,~j $ states, as shown in Eq. (\ref{eq:AmpJ012}). They are obtained in  \cite{branz} and  \cite{raquel}. The $t$ matrices are obtained by means of the Bethe Salpeter equations, and the potentials $V^{I,~j}$ for all the cases that we have in Eq. (\ref{eq:AmpJ012}) are given in Tables XI, XII of Ref. \cite{branz}. 
To these potentials we add an imaginary part to account for the decay into $ D \bar{K} $ for $ I=0, ~ j( D^*  \bar{K}^{*} )=0,~2 $ and into   $   D^*  \bar{K}$ for $ I=0, ~ j( D^*  \bar{K}^{*} )=1,~2 $. For the decay into $ D^*  \bar{K} $ we find \cite{raquel} 
\begin{eqnarray}
i ~Im~V^{(D^{*} \bar{K} )}= -i~A_{j}~\dfrac{1}{8 \pi} (G^{'} g ~m_{D^*} )^{2} q^{5} \left( \dfrac{1}{(\omega_{K}(q)-p^{0}_{2})^{2}-\vec{q}^{2}-m_{\pi}^{2}} \right)^{2} \dfrac{1}{\sqrt{s}} F'^{~4}(q)
\label{eq:boxAnamolous} 
\end{eqnarray}
where we have used  $q=\frac{\lambda^{1/2}(s',m^2_{D^*},m^2_{\bar{K}})}{2\sqrt{s'}}$, $p^{0}_{2}=\dfrac{s'+m_{\bar{K}^{*}}^{2}- m_{D^*}^{2}}{2\sqrt{s'}}$, $ F'(q)=e^{((w_K-p_2^0)^2-\vec{q}\,^2)/\Lambda^2}\ $ , $G'=\frac{3g'^{2}}{4\pi^2f};g'=-\frac{G_Vm_\rho}{\sqrt{2}f^2}$, $G_V\simeq 55$ MeV, $ f=93 $ MeV. As in \cite{raquel} we take $ A_{j}=3/2 $ for $ j=1 $ and  $ A_{j}=9/10 $ for $ j=2 $. 
The coupling $ G^{'} $ in Eq. (\ref{eq:boxAnamolous}) is the one appearing in the anomalous Lagrangian for the vector-vector-pseudoscalar vertex 
\begin{eqnarray}
   {\cal{L}}_{VVP} &=& \dfrac{ G^{'}}{\sqrt{2}} \epsilon^{\mu\nu\alpha\beta} \,\langle \partial_\mu V_\nu \partial_\alpha V_\beta  P \rangle
\label{eq:Lagrangian_VVP}   
\end{eqnarray}
where $V$, $ P $ are the $ q \bar{q} $ matrices written in terms of vectors and pseudoscalar respectively (see Appendix A). The expressions for $ G^{'} $ used above come from Ref. \cite{bramon}, and $ G_{V} $, $ f $ are couplings appearing in chiral Lagrangians. The function $ F'(q) $ is a form factor used in the box diagrams from where $Im V$ is obtained, which we take from Ref. \cite{navarra}.

The decay into  $ D \bar{K} $ is studied in \cite{branz} and a long formula is obtained for the box diagrams containing  $ D \bar{K} $ intermediate state.  The imaginary part alone is easier to evaluate and we sketch its derivation in Appendix A. We consider the imaginary part only for the $I=0$, $D^* \bar{K}^{*} $ states. The reason is that only the $I=0$ states of $ D^*  \bar{K}^{*}  $ are bound. The $  D^*  \bar{K}^{*} $ interaction in $I=1$ is repulsive. The amplitudes of $I=1$ are then small compared to those of $I=0$ which develop poles in the bound region, and then, neglecting a small imaginary part of the $I=1$ amplitudes has a negligible effect in the final results concerning the three body bound states. 

The result for $I=0$ of the imaginary part for decay into $ D \bar{K} $ is given by 
\begin{eqnarray}
i ~Im~V^{(D \bar{K} )}= -i~F_{j}~g^{4}~\dfrac{9}{15}~ \dfrac{1}{2 \pi} ~\dfrac{1}{\sqrt{s'}}~ q^{5} \left( \dfrac{1}{(\omega_{K}(q)-p^{0}_{2})^{2}-\vec{q}^{2}-m_{\pi}^{2}} \right)^{2}  F'^{~4}(q) F_{HQ}
\label{eq:box} 
\end{eqnarray}
where $q=\frac{\lambda^{1/2}(s',m^2_{D},m^2_{\bar{K}})}{2\sqrt{s'}}$, $F_{HQ} = \left(  \dfrac{m_{D^*}}{m_{\bar{K}^{*}}} \right)^{2}$  and  $ F_{j}=5 $ for $ j=0 $ and  $ F_{j}=2$ for $ j=2 $. 
Then the $ t_{1} $ matrix is obtained as 
\begin{equation}  
  t_{1} =\dfrac{1}{V^{-1}-G_{D^*  \bar{K}^{*} }  }, 
  \label{eq:t1tilde} 
  \end{equation}
and to get the same results as in \cite{raquel}, $ G_{D^*  \bar{K}^{*} } $ is regularized with dimensional regularization with the same input as in \cite{raquel}. In the $ G_{D^*  \bar{K}^{*} } $  function we also perform a convolution to account for the width of the $ \bar{K}^{*} $ as done in \cite{branz,raquel}.

 \section{\label{sec:results} Results}
 \begin{center}
  \begin{figure}
  \resizebox{0.8\textwidth}{!}{
  \includegraphics{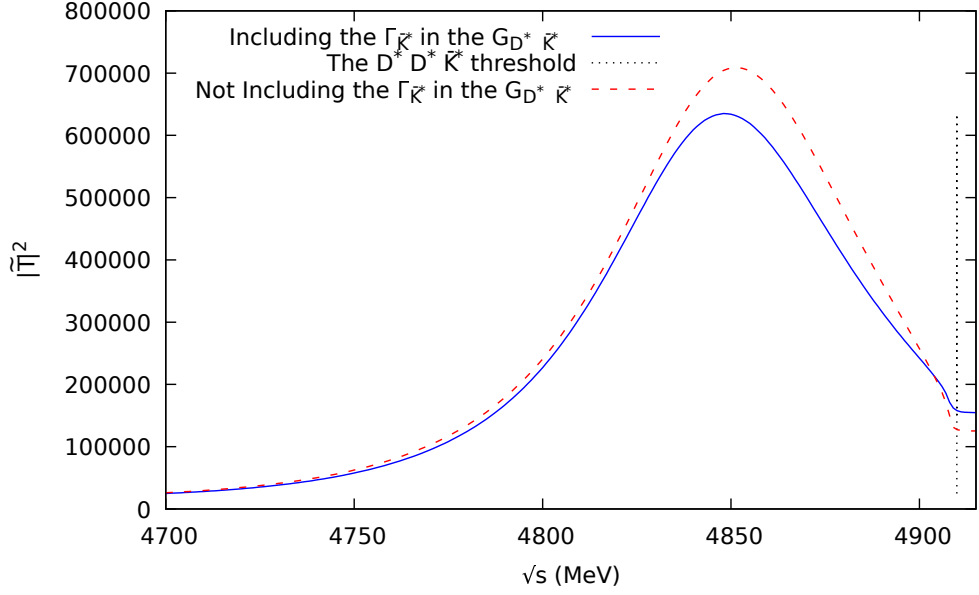}  }
  \caption{(Color in online) Modulus squared of the $ D^*  D^* \bar{K}^{*}$ scattering amplitude with total spin $J=0$. The dashed curve ignores the $\bar{K}^{*}$ width and the solid one accounts for it via a convolution of the $ G_{D^*  \bar{K}^{*} } $  function. The dotted vertical line indicates the $ D^*  D^* \bar{K}^{*}$ threshold.}
  \label{fig:t1}
  \end{figure}
  \end{center} 
 In Fig. \ref{fig:t1} we show the results for $ \mid \tilde{T} \mid^{2} $ as a function of the energy of the   $ D^*  D^* \bar{K}^{*}$ system for $J=0$. We find a clear peak around $4845$ MeV, about $61$ MeV below the  $ D^*  D^* \bar{K}^{*}$ threshold.
To understand this binding we can recall that the  $ D^*  D^* $ state is bound by about $ 4-6 $ MeV, while the $D^* \bar{K}^{*}$ state, corresponding to the  $ X_{0} (2866) $ ($ X_{0} (2900) $ officially), is bound by about $30$ MeV.  
The interaction of $\bar{K}^{*}$  with two  $ D^* $ would give rise to a binding about twice as big as the one of $ D^* \bar{K}^{*}$, and this accounts for the $60$ MeV binding of the $ D^*  D^* \bar{K}^{*}$ state. The width considering the convolution for the $\bar{K}^{*}$ width is about $80$ MeV,  while without the convolution it is about $77$ MeV. We see that the consideration of the $ \bar{K}^* $ width by means of a convolution of the $ G_{D^* \bar{K}^{*}} $ function reduces the strength of $ \mid \tilde{T} \mid^{2}$ and increases a bit the width, but barely changes the mass. As we have seen in Eq. (\ref{eq:AmpJ012}), in this case we have only $ j_{ D^* \bar{K}^{*}} =1 $ and the width comes from the imaginary part of $ t^{I=0,~j=1} $, which has its source in the decay to $ D^* \bar{K} $ (leaving apart the $ \bar{K}^* $ width effect).

\begin{center}
  \begin{figure}
  \resizebox{0.8\textwidth}{!}{
  \includegraphics{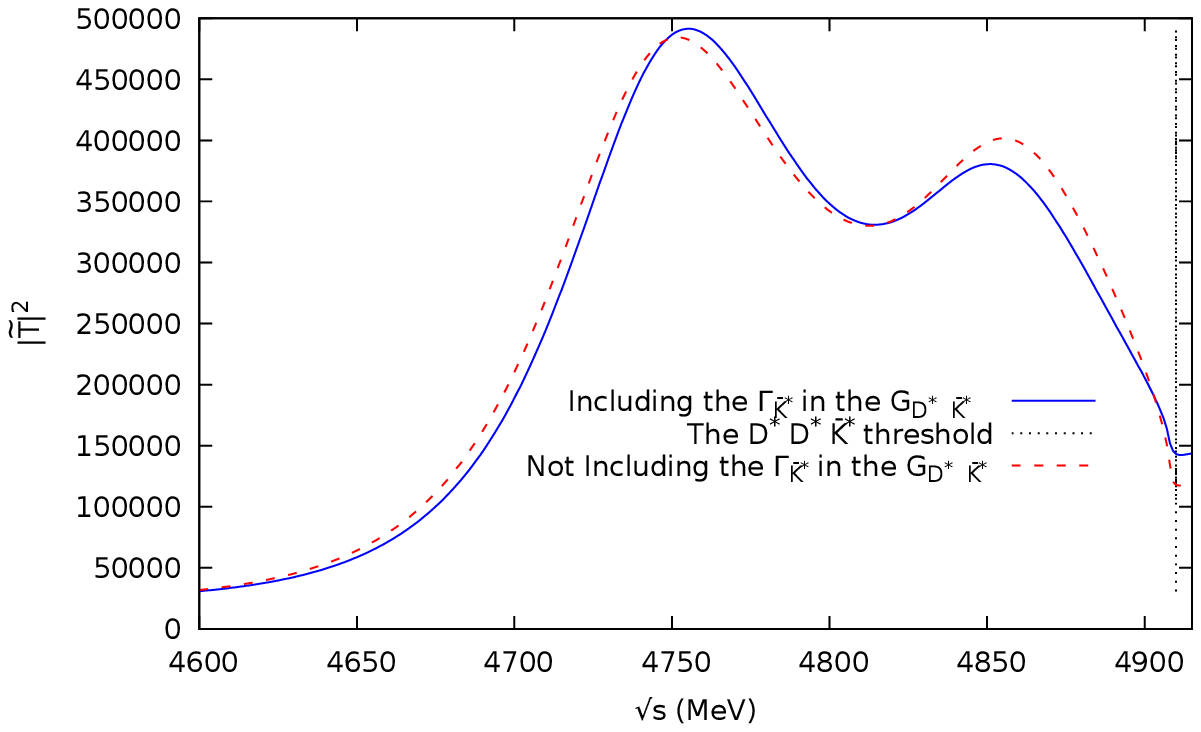}  }
  \caption{(Color in online) Modulus squared of the $ D^*  D^* \bar{K}^{*}$ scattering amplitude with total spin $J=1$. Same meaning of the lines as Fig. \ref{fig:t1}.}
   \vspace{10 mm}
  \label{fig:t2}
  \end{figure}
  \end{center}

\begin{center}
  \begin{figure}
  \resizebox{0.8\textwidth}{!}{
  \includegraphics{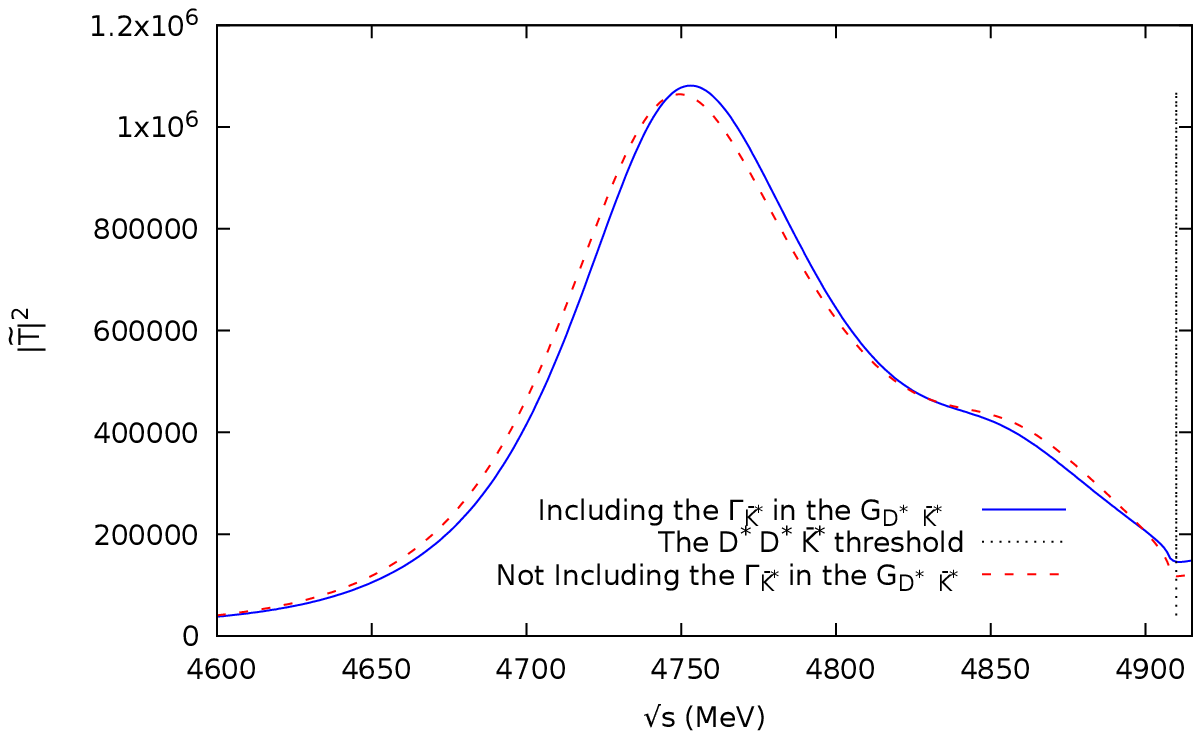}  }
  \caption{(Color in online) Modulus squared of the  $ D^*  D^* \bar{K}^{*}$ scattering amplitude with total spin $J=2$. Same meaning of the lines as Fig. \ref{fig:t1}.}
  \label{fig:t3}
  \end{figure}
  \end{center}

 In Fig. \ref{fig:t2} we show the same picture for total spin $J=1$. Interestingly, we see now two peaks, indicating two states. In this case it is easy to trace the origin of the peaks. As we can see in Eq. (\ref{eq:AmpJ012}), for $J=1$ we have now contributions from $j=0,1,2$. We can see in Table 5 of \cite{raquel} that the $ 1^+ $ and  $ 0^+ $ states have about the same mass, $2861$ MeV and $2866$ MeV respectively, but the $2^+$ state is more bound, with a mass of $2775$ MeV. It is then clear that the first peak (higher energy), that we call state I, is due to $ t^{I=0,~j=0,1} $ while the second peak,  that we call state II, is due to $ t^{I=0,~j=2} $. The effect of the convolution due to the $K^{*}$ width is small, as in the case of $J=0$. 
 In Fig. \ref{fig:t3} we see a similar pattern for the case of total $J=2$. From Eq. (\ref{eq:AmpJ012}) one can also see that now one has $j=1,2$ contribution. The first peak (showing as a shoulder in the figure)  comes from the $ t^{I=0,~j=1} $ amplitude while the second peak is due to $ t^{I=0,~j=2} $ . 
 Thus, in total we find 5 states which we summarize in Table \ref{tab:tab1}. There we also show the main decay channel expected, based on the source of imaginary part of the $t_{1}$ amplitudes involved in the peak. The binding energies range from $56$ MeV to $158$ MeV and the widths 
range from $80$ MeV to $100$ MeV. 
 \begin{table*}[tbh!]
\renewcommand\arraystretch{0.5}
\centering
\caption{ Calculated  Mass ($M$), binding ($B$), width ($\Gamma$) (including the effects of the $ \bar{K}^{*}$ width) of the molecular  $ D^*  D^* \bar{K}^{*}$ states obtained. The binding energy $B$ is measured with respect to the threshold energy $ 2 m_{D^*}+ m_{K^*} $. We also show the main decay mode expected for each one of states.}\label{tab:tab1}
 \begin{tabular*}{1.00\textwidth}{@{\extracolsep{\fill}}   c c c c c c}
 \hline
 \hline
 \\
  $J$&&$M[\mathrm{MeV}]$& $B[\mathrm{MeV}]$ & $\Gamma[\mathrm{MeV}]$& Main decay mode \\
  \hline
  \\
  $0$ &(State I )& $4845$ & $61$ &$80$ & $D^* D^* \bar{K}$ \\
  \\
  $1$ &(State I )& $4850$ &  $56$ &$94$ & $D^* D \bar{K}$, $D^* D^*\bar{K}$ \\
  \\
  $1$ &(State II) &$4754$ & $152$ & $100$&$D^* D \bar{K}$, $D^* D^*\bar{K}$\\
  \\
  $2$ &(State I )& $4840$ & $66$ & $85$ & $D^* D^*\bar{K}$ \\
  \\
  $2$ &(State II )& $4755$ & $151$ & $100$ & $D^* D \bar{K}$, $D^* D^*\bar{K}$ \\
  \\
  \hline
  \hline
 \end{tabular*}
\end{table*} 

It is interesting to note that the strength of the peak of the two states for $J=1$ are similar, while for $J=2$ the strength of the second state is bigger than that of the first one. We can see the reason in Eq. (\ref{eq:AmpJ012}). As we mentioned, the reason that the state II is more bound than the state I, is due to the contribution of the $I=0,~~j=2$ amplitude of $  D^* \bar{K}^{*}$. This reflects the fact that the 
$  D^* \bar{K}^{*}$  state of $I=0,~~j=2$ is more bound than those of $I=0,~~j=0,~ 1$. In Eq. (\ref{eq:AmpJ012}) we see that the weight of $t^{I=0,~~j=2}$ for $J=2$ is $3/16$ while for $J=1$ it is $5/48$, $1.8$ times smaller. On the other hand the weight of $t^{I=0,~~j=1}$ is the same for $J=1$ and $J=2$, and in the case of $J=1$ there is also contribution of $t^{I=0,~~j=0}$ which is absent in $J=2$. 

We address here a different point. By looking at Table \ref{tab:tab1} we can see that the states I for $J=0,1,2$, have similar energies, and so is the case for the two states II with  $J=1$ or $J=2$. It would be good to determine the spin experimentally, a task always difficult but addressed successfully in recent experimental analyses. Partial wave analysis in meson or photon nucleon experiments is performed by different groups and spin and parity of resonances is determined \cite{bonn,deborah,alfred}. Concerning states in the charm sector, LHCb, BelleII and BESIII use also different methods of partial wave analysis to determine spin and parity of the states \cite{lhcb92,belle,bes}. One method that also proved efficient for determining the spin of particles is the use of the moments of angular distributions, where cross sections (not the unknown amplitudes) are projected with spherical harmonics and useful relationships are obtained that help in the determination of the spin \cite{lhcbmom,lhcbtwo,miguel2,mathieu,bayarmom}. 

Another issue we would like to address here is the shape of the wave function that we obtained. We do that for the $J=0$ state. As we have discussed along the paper, we have a $ \bar{K}^{*}$ orbiting around the cluster of $D^* D^*$. The  $ \bar{K}^{*}$ will orbit around one $D^*$ and sometimes around the other $D^*$. In this picture we can have the $ \bar{K}^{*}$ distribution given by 
\begin{equation}  
 \mid \Psi (r^{\prime}_{3})\mid^{2} =\int d^{3}r_{1} d^{3}r_{2} \left( \mid \phi (\vec{r}_{31})\mid^{2} +  \mid \phi (\vec{r}_{32})\mid^{2}  \right)  \mid  \phi^{\prime} (\vec{r}_{12})\mid^{2} \delta^{3} ( m_{D^*} \vec{r}_{1} + m_{D^*} \vec{r}_{2} + m_{\bar{K}^{*}} \vec{r}_{3})
  \label{eq:distribution} 
\end{equation}
where $ \vec{r}_{1} $,  $ \vec{r}_{2} $,  $ \vec{r}_{3} $ are the coordinates of the two  $D^*$ and $\bar{K}^{*}$ respectively and we refer $ \vec{r}^{~\prime}_{3} $ for the  $\bar{K}^{*}$ with respect to the center of mass of the three body system. The wave functions $ \phi $ stand for the $  D^* \bar{K}^{*}$  systems and $ \phi' $ for the  $D^* D^*$ one, and the arguments are given by  $ \vec{r}_{31}= \vec{r}_{3} - \vec{r}_{1}$,  $ \vec{r}_{32}= \vec{r}_{3} - \vec{r}_{2}$,  $ \vec{r}_{12}= \vec{r}_{1} - \vec{r}_{2}$. The wave functions are taken as Eq. (\ref{Eq:Psip}) in momentum space and by means of a Fourier transformation the wave functions  in coordinate space are obtained (See Appendix B). 

  \begin{center}
  \begin{figure}
  \resizebox{0.8\textwidth}{!}{
  \includegraphics{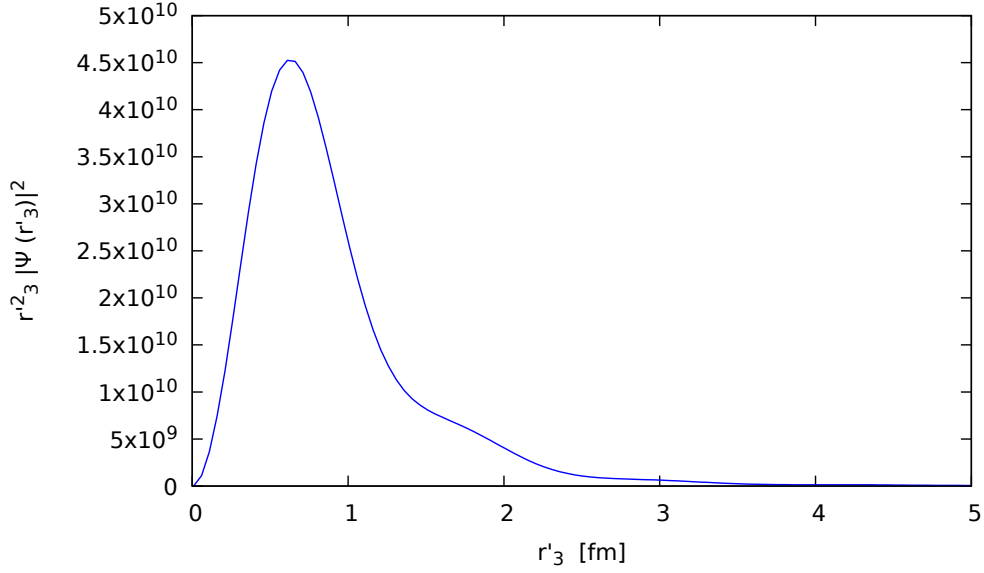}  }
  \caption{(Color in online) The distribution of  $ r'^{2}_{3} \mid \Psi (r'_{3})\mid^{2} $ for the $ \bar{K}^{*}$  in the $ D^*  D^* \bar{K}^{*}$ system at rest. Arbitrary units in y-axis (we take $g_R=1$ in Eq. (\ref{Eq:Psip})).}
  \label{fig:psi}
  \end{figure}
  \end{center}

In Figure \ref{fig:psi} we show the distribution of the $ \bar{K}^{*}$  wave function by means of $ r'^{2}_{3} \mid \Psi (r'_{3})\mid^{2} $ with $ r'_{3}$
 the $ \bar{K}^{*}$ coordinate with respect to the center of mass of the three body system. The  $ r'_{3}$ distribution is plotted without normalization to show the spatial distribution. 
 
As we see, the $ r'_{3}$  distribution peaks around $0.7$ fm. One can evaluate the mean square radius from there and one obtains $ \langle r^{2} \rangle \simeq 1  $ fm  which is bigger than the mean square radius of the proton, $0.84$ fm, and smaller than the one of the deuteron, $2.1$ fm, which is lightly bound and governed by the range of the one pion exchange, while here the range of the interaction is shorter (vector exchange) and the molecule is much more bound. 
  
 \section{Conclusions}

We have conducted a search for possible bound states of the three body system $ D^*  D^* \bar{K}^{*}$. For this we have relied upon the FCA to the Faddeev equations which has proved reliable in the study of related systems compared to the other methods like the variational one using the Gaussian expansion method. For the study of this system we have chosen as the cluster the  $ D^*  D^* $, which appears bound in $I=0$ and $J^{P}=1^{+}$, and the $\bar{K}^{*}$  collides with it repeatedly. The FCA method has the virtue of allowing one to trace the structures obtained to the scattering amplitudes of $  D^* \bar{K}^{*}$. This latter system has bound states in $j=0,1,2$, with the $j=0,1$ states very close in energy and the $j=2$ state more bound. In the $ D^*  D^* \bar{K}^{*}$ system we can have total spin  $J=0,1,2$, and in each of these spins the $D^* \bar{K}^{*}$ amplitude appears with different combinations of isospin and spin. As a consequence, we obtain two peaks in the three-body partition functions which are tied to the $j=0,1$ and $j=2$  of $ D^* \bar{K}^{*}$, with the exception of $J=0$, where the $  D^* \bar{K}^{*}$system can only be in $j=1$, and here we only get one peak. 

In total we obtained $5$ three-body states, one for  $J=0$, two for  $J=1$ and two for  $J=2$, and we evaluate the mass and width of these states. The states obtained, stemming from the cluster of  $ D^*  D^* $ with $I=0$ and $J^{P}=1^{+}$, have total isospin $I=1/2 $, negative parity and spin $J=0,1,2$. The widths range from $80$ MeV to $100$ MeV and the bindings range from $56$ MeV to $151$ MeV. We have also shown which is the main decay channel of each particular state to facilitate its experimental search. We hope the present work stimulates future searches for these super exotic states containing $ccs$ quarks. As mentioned in Ref. \cite{gengminirev}, since the flavor is conserved in the strong interactions, we can expect new states made of many mesons, relatively stable, which cannot decay to mesons with smaller meson number. This would make them similar to nuclei where the baryon number conservation is responsible for the elements of the periodic table, giving rise to a new periodic table of mesons. In coincidence with the point made in Ref. \cite{gengminirev} we think that we are just at the beginning of this new era.

  \section*{Acknowledgments}  
  
  The work of N. I. was partly supported by JSPS KAKENHI Grant Number JP19K14709.
This work is partly supported by the Spanish Ministerio de
Economía y Competitividad (MINECO) and European FEDER funds
under Contract No. PID2020-112777GB-I00, and by Generalitat Va-
lenciana under contract PROMETEO/2020/023. This project has re-
ceived funding from the European Union Horizon 2020 research
and innovation programme under the program H2020-INFRAIA-
2018-1, grant agreement No. 824093 of the STRONG-2020 project.

\section*{Appendix A}  

\begin{center}
  \begin{figure}
  \resizebox{1.\textwidth}{!}{
  \includegraphics{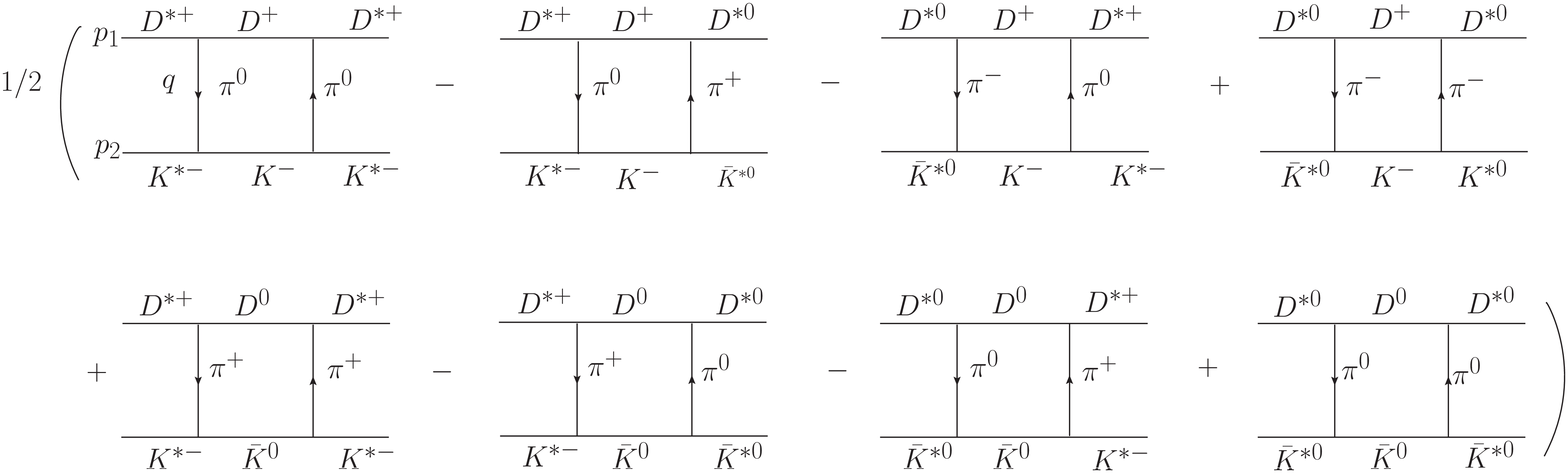}  }
  \caption{ Box diagrams accounting for the decay of $D^{*}\bar{K}^{*}$ in $I=0$ into $  D \bar{K}  $.}
  \label{fig:box}
  \end{figure}
  \end{center}

With the $D^*$, $\bar{K}^{*}$   doublets $(D^{*+}, - D^{*0})$ $(\bar{K}^{*0}, - K^{*-})$, the $I=0$ wave function is given by 
\begin{eqnarray}
-\frac{1}{\sqrt{2}} ( D^{*+} K^{*-}- D^{*0} \bar{K}^{*0}) \nonumber 
\end{eqnarray}
and the box diagrams that account for $  D \bar{K}  $ decay are given in Fig. \ref{fig:box}

The dynamics comes from the vector $ \rightarrow $ pseudoscalar-pseudoscalar vertex, $VPP$ given by the Lagrangian 
\begin{eqnarray}
{\cal L}=-ig \langle [P,\partial_\mu P]V^\mu \rangle, ~~g=\dfrac{M_V}{2f} ~~(M_V=800 ~ \text{MeV}, ~~f=93 ~\text{MeV})\nonumber
\label{eq:lag}
\end{eqnarray}
where $P$, $V$ are $q \bar{q}$ matrices written in terms of pseudoscalar and vector mesons. 
\begin{equation}\label{Psematrix}
P = \begin{pmatrix}
\frac{1}{\sqrt{2}}\pi^0 + \frac{1}{\sqrt{3}} \eta + \frac{1}{\sqrt{6}}\eta' & \pi^+ & K^+ & \bar{D}^0 \\
 \pi^- & -\frac{1}{\sqrt{2}}\pi^0 + \frac{1}{\sqrt{3}} \eta + \frac{1}{\sqrt{6}}\eta' & K^0 & D^- \\
 K^- & \bar{K}^0 & -\frac{1}{\sqrt{3}} \eta + \sqrt{\frac{2}{3}}\eta' & D_s^- \\
D^0  & D^+ & D_s^+ & \eta_c
\end{pmatrix},
\end{equation}
\begin{equation}\label{Vecmatrix}
V = \begin{pmatrix}
 \frac{1}{\sqrt{2}}\rho^0 + \frac{1}{\sqrt{2}} \omega & \rho^+ & K^{* +} & \bar{D}^{* 0} \\
 \rho^- & -\frac{1}{\sqrt{2}}\rho^0 + \frac{1}{\sqrt{2}} \omega  & K^{* 0} & D^{* -} \\
 K^{* -} & \bar{K}^{* 0}  & \phi & D_s^{* -} \\
 D^{* 0} & D^{* +} & D_s^{* +} & J/\psi
\end{pmatrix},
\end{equation}
To simplify the calculation we evaluate the diagrams at threshold of $D^* \bar{K}^{*}$. The external three momenta are zero and $\vec{q}$ is the running variable in the loop.  In this limit the $\epsilon^{0}$ component of the external vectors is zero and all the vertices are of the type of $\vec{\epsilon} \cdot \vec{q}$. Since one is concerned about the imaginary part, one realizes that the pions will be off shell because $D^* \rightarrow D \pi$ is possible but then $\bar{K}^* +\pi \rightarrow \bar{K}$ does not go. 
The loop function is then given by 

 \begin{equation}
i\int\frac{d^4q}{(2\pi)^4} \left( \dfrac{1}{{q^0}^2-\vec{q}\,^2-m_{\pi}^2}\right)^{2}   \frac{1}{(p_{1}-q)^2-m_{D}^2+i\epsilon} \frac{1}{(p_{2}+q)^2-m_{K}^2+i\epsilon},\nonumber
\end{equation}
and we can perform the $q^0$ integration analytically using Cauchy's integration. Since we are concerned only about the imaginary part coming from the $  D \bar{K}  $ placed on shell, we consider only the poles of the $D$ and $\bar{K}  $ propagators and obtain
 \begin{equation}
 \int\frac{d^3q}{(2\pi)^3}  \left( \dfrac{1}{(w_{K}-{p_2}^{0})^2-\vec{q}\,^2-m_{\pi}^2}\right)^{2} \dfrac{1}{2 w_{K}} \dfrac{1}{2 w_{D}}
 \dfrac{1}{{p_1}^{0}+{p_2}^{0}- w_{K} - w_{D} +i\epsilon}. \nonumber
\end{equation}
We have integrals of type 

 \begin{eqnarray}
 \epsilon^{l} \epsilon^{m} \epsilon^{r} \epsilon^{s} \int\frac{d^3q}{(2\pi)^3}  f(\vec{q}\,^2)q^l q^m q^{r}q^{s} = \frac{1}{15}\int \frac{d^3q}{(2\pi)^3}f(\vec{q}\,^2)\vec{q}\,^4
 (\delta_{lm}\delta_{rs}+\delta_{lr}\delta_{ms}+\delta_{ls}\delta_{mr})   \epsilon^{l} \epsilon^{m} \epsilon^{r} \epsilon^{s}, \nonumber
\end{eqnarray}
and considering the $VV$ spin projectors in $ j=0, 1, 2 $ \cite{rhorho}
\begin{eqnarray}
&&{\cal P}^{(0)}= \frac{1}{3}\eps^l \eps^l \eps^r \eps^r \nonumber\\
&&{\cal P}^{(1)}=\frac{1}{2}(\eps^l\eps^m \eps^l \eps^m-\eps^l\eps^m \eps^m \eps^l)\nonumber\\
&&{\cal P}^{(2)}= \frac{1}{2}(\eps^l\eps^m \eps^l \eps^m + \eps^l\eps^m \eps^m \eps^l )-\frac{1}{3} \eps^l \eps^l \epsilon^m \epsilon^m \nonumber
\end{eqnarray}
and the appropriate factors in the vertices, we obtain from the sum of all the diagrams 
\begin{eqnarray}
i ~Im~V^{(D \bar{K} )}= -i~F_{j}~g^{4}~\dfrac{9}{15}~ \dfrac{1}{2 \pi} ~\dfrac{1}{\sqrt{s'}}~ q^{5} \left( \dfrac{1}{(\omega_{K}(q)-p^{0}_{2})^{2}-\vec{q}^{2}-m_{\pi}^{2}} \right)^{2}  F'^{~4}(q) F_{HQ},
\label{eq:box} 
\end{eqnarray}
where we have also included a factor demanded by heavy quark symmetry $F_{HQ}$ to get the coupling $VPP$ in the case of heavy vectors \cite{liangxiao} and used $q=\frac{\lambda^{1/2}(s',m^2_{D},m^2_{\bar{K}})}{2\sqrt{s'}}$, $F_{HQ} = \left(  \dfrac{m_{D^*}}{m_{\bar{K}^{*}}} \right)^{2}$, $p^{0}_{2}=\dfrac{s'+m_{\bar{K}^{*}}^{2}- m_{D^*}^{2}}{2\sqrt{s'}}$, $ F'(q)=e^{((w_K-p_2^0)^2-\vec{q}\,^2)/\Lambda^2}\ $.

Further details can be seen in the evaluation of the width of the analogous $ \bar{B}^{*} \bar{K}^{*} $ state in section D of Ref. \cite{luisBK}.

\section*{Appendix B: Wave function of the  $ D^*  D^* \bar{K}^{*}$ Molecule}  
\begin{center}
  \begin{figure}
  \resizebox{0.4\textwidth}{!}{
  \includegraphics{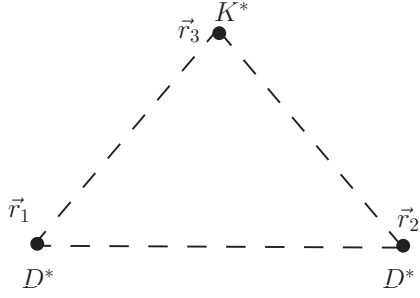}  }
  \caption{ Description of the coordinates of the $ D^*  D^* \bar{K}^{*}$ system.}
  \label{fig:wave}
  \end{figure}
  \end{center}

We have the situation as in Fig. \ref{fig:wave} 
The wave functions for the $\bar{K}^{*}$ relative to the center of mass of the $ D^*  D^* \bar{K}^{*}$ system is given by Eq. (\ref{eq:distribution}). We need the wave functions $\phi (\vec{r}_{31})$,  $\phi (\vec{r}_{32})$, $ \phi^{\prime} (\vec{r}_{12}) $. The wave functions are given in momentum space by Eq. (\ref{Eq:Psip}) $ \langle \vec{q} \mid \phi \rangle  \equiv \phi (\vec{q})$. To obtain the wave function in coordinate space we write:

\begin{equation}  
 \phi (\vec{r}) \equiv \langle \vec{r} \mid \phi \rangle = \int \dfrac{d^{3}q}{(2 \pi )^{3/2}} e^{i \vec{q}\cdot \vec{r}}  \phi (\vec{q})
  \label{eq:phi} 
\end{equation}

\begin{equation}  
 e^{i \vec{q}\cdot \vec{r}} = 4 \pi \sum_{\ell} i^{\ell} j_{\ell}(qr) \sum_{m} Y^{*}_{\ell m} ( \hat{q}) Y_{\ell m} ( \hat{r})
  \label{eq:eps} 
\end{equation}

Performing the $ d\Omega(\hat{r}) $ integration we obtain 

\begin{eqnarray} 
 \phi (\vec{r}) & =& \int \dfrac{{q^2}dq}{(2 \pi )^{3/2}} 4 \pi  \sqrt{4 \pi } j_{0}(qr) Y_{00} ( \hat{x}) \phi (\vec{q})\\ \nonumber
 &=& \dfrac{2 \pi}{(2 \pi )^{3/2}} g_R ~ \frac{2}{r} \int^{q_{max}}_{0} q~dq \dfrac{sin(qr)}{E-\omega_{1}(q)-\omega_{2}(q)}
  \label{eq:phir} 
\end{eqnarray}

Using the $ \delta^{3} ( ~) $ function in Eq. (\ref{eq:distribution}) we can eliminate $\vec{r}_{2}$ and have only the integral over $\vec{r}_{1}$, which requires two integrals, one over the modulus of $\vec{r}_{1}$ and the other one over the angle between $\vec{r}_{1}$ and $\vec{r}_{3}$. Altogether we need three integrals to evaluate $ \mid \Psi (r^{\prime}_{3})\mid^{2} $ of Eq. (\ref{eq:distribution}).

 As mentioned before, to obtain the binding of the $ T_{cc} $, the value $ q_{max}= 420$ MeV was used, and the same value is used to construct the wave function of the $ D^*  D^*$ cluster. On the other hand in Ref. \cite{raquel} the loops were regularized with dimensional regularization. Since here we need the equivalent $ q_{max} $, we have seen which value is needed to obtain the same binding of the $D^* \bar{K}^{*}$ molecule, which is found for $ q_{max}= 1050$ MeV (see also Ref. [82]) that we use here to construct the $  D^* \bar{K}^*  $ wave function.

\end{document}